\documentclass[aps,groupedaddress]{revtex4}

\begin{document}

\title{Coherent State Path Integral for Bloch Particle}
\author{Junya Shibata}
\affiliation{Department of Physics, Osaka University, 
Toyonaka, Osaka 560-0043, Japan}
\author{Komajiro Niizeki}
\affiliation{Department of Physics, Tohoku University, 
Sendai 980-8578, Japan}
\date{\today}

%\begin{center}
%
%{\Large\bf {\bf Coherent State Path Integral for Bloch Particle}}
%
%\par\bigskip
%
%Junya Shibata and Komajiro Niizeki
%
%\par\bigskip
%
%{\it Department of Physics, Osaka University, 
%Toyonaka, Osaka 560-0043, Japan,\\
%Department of Physics, Tohoku University, 
%Sendai 980-8578, Japan}
%
%\end{center}

%%%%%%%%%%%%%%%%%%%%%%%%%%%%%%%%%%%%%%%%%%

\begin{abstract}
We construct a coherent state path integral formalism for the
one-dimensional Bloch particle 
within the single band model. 
The transition amplitude between two coherent states is 
a sum of transition amplitudes with different 
winding numbers on the two-dimensional phase space which has the same
topology as that of the cylinder. 
Appearance of the winding number is due to the periodicity of the
quasi-momentum of the Bloch particle. 
Our formalism is successfully applied to a semiclassical motion of 
the Bloch particle under a 
uniform electric field.
The wave packet exhibits not only the Bloch oscillation but also a similar
breathing to the one for the squeezed state of a harmonic oscillator.
%action approximation. 
%We also present the arbitrary initial condition 
%for the time evolution of Bloch particle in a 
%uniform along-axis electric field on the basis of stationary 
%action approximation. 
\end{abstract}

\maketitle

%%%%%%%%%%%%%%%%%%%%%%%%%%%%%%%%%%%%%%%%%%
\section{Introduction}
%%%%%%%%%%%%%%%%%%%%%%%%%%%%%%%%%%%%%%%%%%

The path integral formalism \cite{Feynman-Hibbs} of quantum mechanics uses
classical paths to
evaluate a quantum-mechanical transition amplitude. This is an advantage of
the formalism over other formulations of quantum mechanics because
intuitive arguments are available on solving separate problems in quantum
mechanics. The path integral formalism is particularly useful when a
physics in a semiclassical regime is considered. In quantum mechanics,
dynamical variables are usually quantized and can be mutually incommutable.
On the other hand, in the classical mechanics, they are continuous and are
mutually commutable. The path integral formalism fully exploits these
properties of the classical mechanics. A typical example is the coherent
state path integral formalism \cite{Klauder}, \cite{Schulman1} for the
quantum spin system 
\cite{Kashiwa-Ohnuki-Suzuki}.

Meanwhile, the motion of a single electron in a crystal lattice is
successfully treated by the single band model \cite{Mermin}, 
in which the relevant
Hilbert space is restricted to the subspace associated with a particular
band. We may call such an electron a Bloch particle (electron)
\cite{Bloch2}. We shall consider, for simplicity, the one-dimensional (1D)
Bloch particle. On account of the restricted Hilbert space, the position
and the momentum of the Bloch particle have important differences from
those of the conventional particle: i) the position is quantized in unit of
the lattice spacing $a$ and ii) the momentum is not a true momentum but a
quasi-momentum which has a periodicity with the period $2\pi\hbar/a$. The
two points, i) and ii), are closely related to each other because the
position and the quasi-momentum are canonically conjugate to each other.
They prevent us from directly applying the path integral formalism for the
conventional particle to quantum mechanics of the Bloch particle. If there
ever exists a consistent path integral formalism for the Bloch particle,
the position of the Bloch particle must be treated as a continuous variable
in the formulation.

The purpose of this paper is to present a coherent state path integral 
formalism for the Bloch particle. 
The theory will be applied to the transition amplitude 
of the coherent state of the Bloch particle 
under a uniform electric field, and famous Bloch oscillations 
\cite{Bloch2}, \cite{Zener} will be
reproduced. 

The paper is organized as follows. 
In Section 2 we construct the coherent state for the Bloch particle. 
This is given in the Wannier representation. 
In Section 3 we compute matrix elements of the kinetic energy 
and the potential energy for the coherent states by use of the Poisson
summation formula. 
In Section 4 we construct the coherent state path integral formalism 
for the Bloch particle. 
%The transition amplitude between two coherent states is a sum of transition 
%amplitudes with different winding numbers. 
In Section 5 evaluates the transition amplitude for the Bloch particle 
under a uniform electric field on the basis of the 
stationary-action approximation. 
%The wave packet through the transition probability exhibits not only 
%the Bloch oscillation but also a similar breathing to the one for 
%the squeezed of a harmonic oscillator. 
In Section 6 we conclude this paper with a summary and a discussion. 

%%%%%%%%%%%%%%%%%%%%%%%%%%%%%%%%%%%%%%%%%% 
\section{Coherent State of Bloch Particle}
%%%%%%%%%%%%%%%%%%%%%%%%%%%%%%%%%%%%%%%%%%

The Wannier states, $|n\rangle, ~n \in {\bf Z}$, form 
an orthonormal complete set of the Hilbert space of the Bloch particle:
\begin{equation}
\sum_{n=-\infty}^{\infty}|n\rangle\langle n|=1.
\label{comp-WS}
\end{equation}
The position operator, $\hat{x}$, is diagonal in the Wannier representation:
\begin{equation}
\hat{x}=\sum_{n=-\infty}^{\infty}|n\rangle x_{n}\langle n |,\qquad
x_{n} := na,
\label{position-WS}
\end{equation}
while the translation operator is off-diagonal:
\begin{equation}
\hat{T}=\sum_{n=-\infty}^{\infty}|n+1\rangle\langle n |, 
\label{trans1-WS}
\end{equation}
or, equivalently, $\hat{T}|n\rangle = |n+1\rangle$.
Then, $\hat{x}$ and $\hat{T}$ satisfy the commutation: 
\begin{equation}
[\hat{x},\hat{T}] = a \hat{T}, \label{comm}
\end{equation}
in which the quantization of $\hat{x}$ is built in. 
The quasi-momentum operator $\hat{p}$ is ill-defined but is related to the
well-defined operator $\hat{T}$ by
\begin{equation}
\hat{T}=\exp \left(-\frac{i}{\hbar}a\hat{p}\right). 
\label{trans2-WS}
\end{equation}
%Note that $\hat{p}$ is ill-defined but $\hat{T}$ is.

The translation operator $\hat{T}$, which is unitary, is diagonal in the
Bloch representation which is based on the Bloch state (the momentum
state), $|p\rangle$. That is, 
\begin{equation}
\hat{T}|p\rangle = \exp \left(-\frac{i}{\hbar}ap\right) |p\rangle.
\label{eigenT}
\end{equation}
The Bloch state is specified by its wave function, 
$\langle n |p\rangle = \sqrt{a/(2\pi\hbar)}\exp (inap/\hbar)$. %####
It is periodic: $|p + 2\pi\hbar/a\rangle = |p\rangle$, so that $p$ is a
variable on a circle (or, equivalently, the Brillouin zone) whose radius is
equal to $\hbar/a$. It is unnormalizable but two different Bloch states are
mutually orthogonal. 
The completeness of Bloch states is represented as %####
\begin{equation}
\int_{\rm Bz}dp\,|p\rangle\langle p|=1, 
\label{comp-BS}
\end{equation}
where the symbol Bz stands for the Brillouin zone.

It is convenient to introduce the angular variable, $\phi := ap/\hbar$,
which is the dimensionless quasi-momentum. 
It is a variable on the unit circle, $S^1$. Its operator version,
$\hat{\phi}$, is related to $\hat{T}$ by
\begin{equation}
\hat{T}=\exp{(-i\hat{\phi})}. 
\label{trans-phi}
\end{equation}
Since $\phi$ has a different scale from that of $p$, we shall change the
normalization for $|p\rangle$ to obtain $|\phi\rangle :=
\sqrt{\hbar/a}\,|p\rangle$ or, equivalently, $\langle n |\phi\rangle = \exp
(in\phi)/\sqrt{2\pi}$. Hence, we may write
\begin{equation}
|\phi\rangle =
\frac{1}{\sqrt{2\pi}}\sum_{n=-\infty}^{\infty}e^{in\phi}|n\rangle, 
\label{BS}
\end{equation}
\begin{equation}
\oint d\phi\,|\phi\rangle\langle\phi|=1. 
\label{comp-BS'}
\end{equation}

Our path integral formalism for the Bloch particle will be established
through an extension of the coherent state path integral formalism for the
conventional particle \cite{Schulman1}. In the latter formalism, we use coherent states, in
each of which the position and the momentum are specified within the
restriction imposed by the Heisenberg uncertainty principle. Therefore, the
coherent state is a minimal packet and is an eigenstate of the
non-hermitian operator: $\hat{z} := \kappa\hat{x} + i\lambda\hat{p}$, where
$\kappa$ and $\lambda$ are positive parameters characterizing the coherent
state. Since $[\hat{z}, ~\hat{z}^\dagger] = \kappa \lambda/2$, $\hat{z}$ is
an unnormalized annihilation operator of a harmonic oscillator, and we
assume that $\kappa$ and $\lambda$ are chosen so that $\hat{z}$ is
dimensionless. Different coherent states are different eigenstates:
$\hat{z}|z\rangle= z|z\rangle$,
where $z :=\kappa x+i\lambda p$ is a complex parameter specifying the
coherent state.
The position and the momentum of the coherent state %####
are not sharply defined 
but their mean values are given by $x$ and $p$, respectively, while their
spreads by 
\begin{equation}
\delta x = \sqrt{\frac{\hbar\lambda}{2\kappa}}, \quad \delta p =
\sqrt{\frac{\hbar\kappa}{2\lambda}},
\end{equation}
which satisfy $\delta x \delta p = \hbar/2$. Note that the ratio
$\kappa/\lambda$ is more important than the two separate parameters
themselves. The complex $z$ plane is isomorphic to the two-dimensional phase
space $(x, ~p)$, and $z$ can be treated as a complex dynamical variable in
the coherent state path integral formalism.

We shall return to the case of the Bloch particle.
A drawback for the case is that $\hat{p}$ is ill-defined. To circumvent it,
we set $\lambda = a/\hbar$ and introduce a new non-hermitian operator by
the formal equation, $\hat{A} := \exp{(-\hat{z})}$, which is well-defined
because
\begin{equation}
\hat{z} = \kappa \hat{x}+i\hat{\phi}. 
\label{z-op}
\end{equation}
The formal expression for $\hat{A}$ can be transformed with a standard
procedure of the operator algebra into the exact one:
\begin{equation}
\hat{A} = \exp\left[-\kappa\left(\hat{x}-\frac{a}{2}\right)\right]\hat{T}. %####
\end{equation}
The coherent state, $|z\rangle$, of the Bloch particle is an eigenstate of
$\hat{A}$:
\begin{equation}
\hat{A}|z\rangle=e^{-z}|z\rangle, \quad z :=\kappa x+i\phi. 
\end{equation}
It is of a vital importance in a later argument that the phase space for
the complex dynamical variable $z$ is the cylinder $\Gamma := {\bf R}\otimes
S^{1}$ or, equivalently, the fundamental strip, $\Sigma := -\pi < \Im z \le
\pi$, on the complex $z$ plane. Remember that $\Gamma$ as well as $S^{1}$
is not simply connected.

If the coherent state is normalized appropriately, it is given in the
Wannier representation as
\begin{equation}
|z\rangle =\sum_{n=-\infty}^{\infty}C_{n}(z)|n\rangle,
\label{CS-WS}
\end{equation}
\begin{equation}
C_{n}(z) := \langle n | z\rangle %####
= N \exp \left[
-\frac{\kappa}{2a}(x_{n}-x)^{2}+in\phi
\right],\qquad N=\left(\frac{\kappa a}{\pi}\right)^{1/4}. 
\label{cn-CS}
\end{equation}
Thus, the coherent state is of a wave packet, whose spread is equal to
$\delta x := \sqrt{a/(2\kappa)}$.

The Bloch representation for the coherent state is readily obtained from
this equation together with (\ref{BS}):
\begin{equation}
\langle \phi' | z\rangle 
= \sum_{n=-\infty}^{\infty}\frac{N}{\sqrt{2\pi}}\exp \left[
-\frac{\kappa}{2a}(x_{n}-x)^{2}+in(\phi - \phi')
\right], 
\label{pn-CS}
\end{equation}
which can be rewritten with the Poisson summation formula into
\begin{equation}
\langle \phi' | z\rangle 
= \sum_{m=-\infty}^{\infty}N'\exp \left[
-\frac{1}{2\kappa a}(\phi' - \phi_{m})^{2}-i\frac{x}{a}(\phi' - \phi_{m})
\right], \qquad N' := (\pi\kappa a)^{-1/4}
%N'=\left(\frac{a}{\pi\kappa\hbar^2}\right)^{-1/4}
\label{pn-CS'}
\end{equation}
with $\phi_{m} := \phi + 2\pi m$, where $m$ is an integral variable to be
called the {\it winding number}. 
The coherent state has a form of a wave packet in the momentum space as
well, and its spread is equal to $\delta \phi := \sqrt{\kappa a/2}$, which
is consistently related to $\delta x$ by the Heisenberg uncertainty
equality.

 Although the left hand side of (\ref{pn-CS'}) is periodic on the variable
$\phi'$, each summand in the r.\ h.\ s.\ (right hand side) is not. 
Hence, the variable $\phi'$ should be regarded in the r.\ h.\ s.\ as a
variable on ${\bf R}$, which is simply connected and is the universally
covering space for $S^{1}$. A similar situation will be realized in other
infinite series of periodic functions to appear.

The inner product between different coherent states 
can be calculated in the Wannier representation as
\begin{equation}
\langle z | z'\rangle = N^{2}\sum_{n=-\infty}^{\infty}
\exp \left[-\kappa a n^2 + n(z^{*}+z')-\frac{\kappa}{2a}
(x^{2}+x ^{'2})
\right], 
\label{inner-CS}
\end{equation}
which is transformed with the Poisson summation formula into
\begin{equation}
\langle z | z'\rangle = \sum_{m=-\infty}^{\infty}\exp\left[
\frac{i}{\hbar}{\cal S}_{{\rm c}}(z,z'_{m})
\right], 
\qquad z'_{m}:= z'+2\pi m, 
\label{inner-CS'}
\end{equation}
where 
\begin{equation}
\frac{i}{\hbar}{\cal S}_{{\rm c}}(\zeta,~\zeta') %####
=\frac{1}{8\kappa a}
\left[2(\zeta^{*}+\zeta')^{2}-(\zeta^{*}+\zeta)^{2}-(\zeta'^{*}+\zeta')^{2}
\right]. %####
\label{Sc}
\end{equation}
Note that ${\cal S}_{{\rm c}}(z,z'_{m})$ depends on $m$ because ${\cal
S}_{{\rm c}}(z,~\zeta)$ is periodic on neither of the two variables.
In particular, we obtain, 
\begin{eqnarray}
\langle z |z\rangle &=& \sum_{m=-\infty}^{\infty}\exp\left[
-\frac{\pi^{2}}{\kappa a}m^{2}+i\frac{\pi x}{a}m
\right] \\
&=&\vartheta_{3}\left(\left.\frac{\pi x}{2a}\right|
\frac{i\pi}{\kappa a}\right), 
\end{eqnarray}
where $\vartheta_{3}(z|\tau)$ is the $\theta$-function 
\cite{Whittaker}
defined by 
\begin{equation}
\vartheta_{3}(z|\tau):=\sum_{n=-\infty}^{\infty}
e^{in^{2}\pi \tau + i2\pi nz},\qquad \Im \tau > 0. 
\end{equation}
Hence, the coherent state $|z\rangle$ is not normalized to unity. In fact,
the r.\ h.\ s\verb/./'s of (\ref{pn-CS}), (\ref{pn-CS'}), (\ref{inner-CS}) and
(\ref{inner-CS'}) are all represented with the third $\theta$-function,
$\vartheta_3$, but with different arguments. The transformation combining
the two expressions for $\langle \phi' | z \rangle$ or $\langle z | z'
\rangle$ is the famous Jacobi's imaginary-number transformation for the
$\theta$-function.
%####

We can consider $C^{*}_{n}(z) = \langle z | n\rangle$ to be the wave
function of the Wannier state $| n\rangle$ in the coherent-state
representation. The wave functions of Wannier states are orthonormal in the
sense represented as
\begin{equation}
\int C_{n}(z)C^{*}_{n'}(z) d\mu(z) = \delta_{n,n'}, \qquad d\mu(z) :=
\frac{dx\,d\phi}{2\pi a}, 
\label{orthonormal2}
\end{equation}
where the integral must be performed over the entire phase space.
The coherent states form an overcomplete set but we can derive from this
result the following resolution of unit: %####
\begin{equation}
\int d\mu(z)\,|z\rangle\langle z| = 1.
\label{comp-CS}
\end{equation}
This is a basic result for the construction of the coherent state path
integral formalism. 

Above discussions show that the coherent state of the Bloch particle has
similar properties to those of the coherent state of the conventional
particle. The former coherent state tends asymptotically to the latter in
the quasi-continuous limit, $a\kappa \ll 1$, where the spread of the packet
in the real space is much larger than $a$, the lattice spacing.

It is important in a later discussion that, for a given $\phi'$, the
summation in the r.\ h.\ s.\ of (\ref{pn-CS'}) is dominated by a single term
if $a\kappa \ll 1$, so that the interference among terms with different
winding numbers vanishes then.

%%%%%%%%%%%%%%%%%%%%%%%%%%%%%%%%%%%%%%%%%%%%%%%%%%
\section{Matrix Elements}
%%%%%%%%%%%%%%%%%%%%%%%%%%%%%%%%%%%%%%%%%%%%%%%%%%

In order to construct the coherent state path integral formalism, 
we need to compute the matrix elements of the Hamiltonian, $\hat{H}$, for 
the coherent states. We assume that $\hat{H}$ is written as 
\begin{equation}
\hat{H} = \hat{{\cal W}} + \hat{{\cal G}}, 
\label{H}
\end{equation}
where $\hat{{\cal W}}$ and $\hat{{\cal G}}$ are the kinetic energy and the
potential energy, respectively, and are commutable with $\hat{T}$ and
$\hat{x}$, respectively.

We begin our calculation with $\hat{{\cal G}}$. %####
We assume it to be represented with an analytic function $g(\zeta)$ %####
as $\hat{{\cal G}}=g(\hat{x})$.
Its matrix element for coherent states 
is calculated in the Wannier representation as
\begin{equation}
\langle z |\hat{{\cal G}}|z'\rangle=
N^{2}\sum_{n=-\infty}^{\infty}g(na)
\exp \left[
-\kappa a n^{2}+ n(z^{*}+z')-\frac{\kappa}{2a}(x^{2}+x^{'2})
\right], 
\end{equation}
which is transformed by use of the Poisson summation formula into:
\begin{equation}
\langle z |\hat{{\cal G}}|z'\rangle=
\sum_{m=-\infty}^{\infty}\exp\left[
\frac{i}{\hbar}{\cal S}_{c}(z,z'_{m})\right]G(z^{*},z'_{m}),
\label{zGz'}
\end{equation}
where
\begin{equation}
G(\zeta,\zeta')=\left(\frac{\kappa a}{\pi}\right)^{1/2}
\int_{-\infty}^{\infty}dt\,
e^{-\kappa t^{2}/a}g\left(
t+\frac{\zeta+\zeta'}{2\pi \kappa}
\right).
\label{Gmtrx}
\end{equation}

Meanwhile, the Bloch state $|\phi \rangle$ is an eigenstate of $\hat{{\cal
W}}$, which is commutable with $\hat{T}$. The eigenvalue $W(\phi)$ is a
periodic function: $W(\phi + 2\pi) = W(\phi)$. Let us write the Fourier
expansion of $W(\phi)$ as
%$\hat{{\cal W}}|\phi \rangle = W(\phi)|\phi \rangle$ 
\begin{equation}
W(\phi)
=\sum_{l=-\infty}^{\infty}V_{l}\,e^{-il\phi}, 
\qquad V_{-l}=V^{*}_{l}.
\label{gene-disp}
\end{equation}
If the potential term is absent, the eigen-energy is specified by $\phi$ as
$E = W(\phi)$, which is
nothing but the dispersion relation for the Bloch particle. A simplest
expression for $W(\phi)$ is 
\begin{equation}
W(\phi) = -V\cos{\phi}.
\label{simp-disp}
\end{equation}
The relevant band width is $2V$ provided that $V > 0$.

For (\ref{gene-disp}), we obtain
\begin{equation}
\hat{{\cal W}}
=\sum_{l=-\infty}^{\infty}V_{l}\,\hat{T}^{l}.
\end{equation}
The matrix elements of $\hat{T}^{l}$ for coherent states can be 
calculated by a similar procedure to the one for $\hat{{\cal G}}$: 
\begin{eqnarray}
\langle z |\hat{T}^{l}|z'\rangle &=& 
N^{2}\sum_{n=-\infty}^{\infty}
\exp\left[
-\kappa a n^{2} + n(z^{*}+z'+l\kappa a) 
-lz'-\frac{\kappa}{2a}(x^{2}+x^{'2})-\frac{\kappa a}{2}l^{2}
\right] \nonumber \\
&=&\sum_{m=-\infty}^{\infty}
\exp\left[
\frac{i}{\hbar}{\cal S}_{{\rm c}}(z,z'_{m})
+\frac{l}{2}(z^{*}-z'_{m})-\frac{\kappa a}{4}l^{2}
\right].
\end{eqnarray}
Thus, 
\begin{equation}
\langle z |\hat{{\cal W}}|z'\rangle
=\sum_{m=-\infty}^{\infty}\exp\left[
\frac{i}{\hbar}{\cal S}_{{\rm c}}(z,z'_{m})\right]W(z^{*},z'_{m}), 
\label{zTz'}
\end{equation}
where 
\begin{equation}
W(\zeta,\zeta')=\sum_{m=-\infty}^{\infty}
e^{-\kappa al^{2}/4}
%\left(
V_{l}\,e^{l(\zeta-\zeta')/2}. 
%+V^{*}_{l}e^{-l(z^{*}-z'_{m})/2}
%\right)
\label{Wzzeta}
\end{equation}

From (\ref{H}), (\ref{zGz'}), and (\ref{zTz'}), we obtain
\begin{equation}
\langle z |\hat{{\cal H}}|z'\rangle
=\sum_{m=-\infty}^{\infty}\exp\left[
\frac{i}{\hbar}{\cal S}_{{\rm c}}(z,z'_{m})\right]H(z^{*},z'_{m}) 
\label{zHz'}
\end{equation}
with
\begin{equation}
H(\zeta,\zeta'):= W(\zeta,\zeta') + G(\zeta,\zeta'). 
\label{c-Hamiltonian}
\end{equation}

%%%%%%%%%%%%%%%%%%%%%%%%%%%%%%%%%%%%%%%%%%
\section{Coherent State Path Integral Formalism}
%%%%%%%%%%%%%%%%%%%%%%%%%%%%%%%%%%%%%%%%%%

Now, we can construct a coherent state path integral formalism for the
Bloch particle. 
We write the transition amplitude between 
the initial state $|z_{{\rm I}}\rangle$ and the final state 
$|z_{{\rm F}}\rangle$ as 
\begin{equation}
{\cal K}(z_{{\rm F}},z_{{\rm I}};T) := \langle z_{{\rm F}}|
\exp{(-iT\hat{H}/\hbar)}|z_{{\rm 
I}}\rangle.
\label{tp1}
\end{equation}
We slice the time interval into $N$ 
identical pieces of length 
$\epsilon := T/N$, and write $\exp{(-iT\hat{H}/\hbar)} =
[\exp{(-i\hat{H}\epsilon/\hbar)}]^N$.
Inserting the resolution of unit, (\ref{comp-CS}), into the %####
relevant site associated with each discrete time yields
\begin{equation}
{\cal K}(z_{{\rm F}},z_{{\rm I}};T) 
=\lim_{N\to\infty}\int
\prod_{j=1}^{N-1}d\mu(z_{j})\prod_{j=1}^{N}
\langle z _{j}|e^{-i\epsilon\hat{H}/\hbar}|z_{j-1}\rangle %####
\label{tp2} 
\end{equation}
with $z_{0} := z_{{\rm I}}$ and $z _{N} := z_{{\rm F}}$. %####
The transition amplitude $\langle z|
\exp{(-i\epsilon\,\hat{H}/\hbar)}|z'\rangle$ can be rewritten to the order
$\epsilon$ as 
\begin{eqnarray}
\langle z |e^{-i\epsilon\,\hat{H}/\hbar} |z'\rangle
&\simeq& \langle z |\left(1-i\epsilon\hat{H}/\hbar\right)|z'\rangle \nonumber \\
&=& \langle z|z'\rangle-\frac{i\epsilon}{\hbar}
\langle z|\hat{H}|z'\rangle
%\label{*}
\end{eqnarray}
Using (\ref{inner-CS'}) and (\ref{zHz'}), %####
we can rewrite this
equality to the order $\epsilon$ as
\begin{equation}
\langle z |e^{-i\epsilon\,\hat{H}/\hbar} |z'\rangle 
\simeq \sum_{m =-\infty}^{\infty}
\exp\left[
\frac{i}{\hbar}
{\cal S}(z,z'_{m})
\right], 
\label{inf-tp}
\end{equation}
with 
\begin{equation}
{\cal S}(\zeta,\zeta'):=
{\cal S}_{{\rm c}}(\zeta,\zeta') - \epsilon\,H(\zeta,\zeta').
\label{ta3}
\end{equation}
Using (\ref{inf-tp}), we can change (\ref{tp2}) into
we obtain 
\begin{equation}
{\cal K}(z_{{\rm F}},z_{{\rm I}};T) = \lim_{N\to\infty}\int
\prod_{j=1}^{N-1}d\mu(z_{j})
\prod_{j=1}^{N}
\sum_{m_{j}=-\infty}^{\infty}
\exp\left[\frac{i}{\hbar}{\cal S}(z_{j},z_{j-1} + 2\pi im_{j})
\right]. 
\label{cspi1}
\end{equation}

Since $H(\zeta,\zeta')$ is composed of two terms, $W(\zeta,\zeta')$ and
$G(\zeta,\zeta')$,
${\cal S}(\zeta,\zeta')$ has three terms. From (\ref{Sc}), (\ref{Gmtrx}) and
(\ref{Wzzeta}), we see that ${\cal S}(z_{j},z_{j-1} + 2\pi im_{j})$ depends
on $m_{j}$ through the two forms: 
\begin{equation}
z^{*}_{j} + z_{j-1} + 2\pi im_{j}, \qquad z^{*}_{j} - z_{j-1} - 2\pi im_{j}. 
\label{two forms}
\end{equation}

We can rewrite (\ref{cspi1}) into a more convenient form if we employ a
procedure 
in a text book \cite{Kashiwa-Ohnuki-Suzuki}, 
in which the authors discuss the Feynman kernel in a periodic system. 
We begin with changing the integer variables, $m_{j}$'s, into others,
$m'_{j}$'s, by
\begin{equation}
m_{j} = m'_{j}-m'_{j-1},\qquad m'_{0} = 0, \qquad j=1,2,\cdots,N. %####
\end{equation}
Then, the sum on $m_{j}$'s is transformed to that on $m'_{j}$'s.
We next change $\phi_{j}$ into 
\begin{equation}
\phi_{j}=\phi'_{j}+2\pi m'_{j}, \qquad j=1,2,\cdots,N-1. 
\end{equation}
Then, two quantities in (\ref{two forms}) are transformed to
\begin{equation}
z^{'*}_{j}+z'_{j-1}, \qquad z^{'*}_{j}-z'_{j-1} - 4\pi im'_{j}, 
\label{primed quantity}
\end{equation}
respectively, where $z'_{j} := \kappa x_{j}+i\phi'_{j}$. Since $z_{j}$ is a
variable on the fundamental strip $\Sigma$, $z'_{j}$ is a variable on the
shifted strip, $\Sigma - 2\pi im'_{j}$. %####
From the considerations made up to
this point, we can conclude that the integration on $\phi_{j}$'s in
(\ref{cspi1}) is transformed as
\begin{equation}
\prod_{j=1}^{N}\sum_{m_{j}=-\infty}^{\infty}\prod_{j=1}^{N-1}
\int_{-\pi}^{\pi}d\phi_{j}
=\sum_{m'_{N}=-\infty}^{\infty}\prod_{j=1}^{N-1}%####
\int_{-\infty}^{\infty}d\phi'_{j}, 
\end{equation}
where the quantity ${\cal S}(z_{j},z_{j-1} + 2\pi im_{j})$ included in the
summand of (\ref{cspi1}) must be replaced simultaneously by ${\cal
S}(z'_{j},z'_{j-1})$. 
Note that the term $-4\pi im'_{j}$ in the second quantity of (\ref{primed
quantity}) does not effect the summand because of the form of
(\ref{Wzzeta}). 
Thus, we arrive at the final expression for the transition amplitude: 
\begin{equation}
{\cal K}(z_{{\rm F}},z_{{\rm I}};T) 
:= \sum_{m=-\infty}^{\infty}{\cal L}(z_{{\rm F}}^{m},z_{{\rm I}};T), \qquad 
z_{{\rm F}}^{m} := z_{{\rm F}}+i2\pi m
\label{summ-amplitude}
\end{equation}
with 
\begin{equation}
{\cal L}(z_{{\rm F}}^{m},z_{{\rm I}};T)
=\lim_{N\to\infty}\prod_{j=1}^{N-1}
\int_{-\infty}^{\infty}\int_{-\infty}^{\infty}
\frac{dx_{j}}{a}\frac{d\phi_{j}}{2\pi}
\exp
\left(\frac{i}{\hbar}{\cal S}[{\bf z}^{*},{\bf z}]\right), %####
\label{T-amplitude}
\end{equation}
where the action ${\cal S}[{\bf z}^{*},{\bf z}]$ is a ``functional" of the
path ${\bf z} := 
(z_{{0}}, ~z_{1}, ~\cdots, ~z_{N})$ with $z _{N} := z_{{\rm F}}^{m}$. 
and is given by
\begin{equation}
{\cal S}[{\bf z}^{*},{\bf z}] = \sum_{j=1}^{N}{\cal S}(z_{j}, z_{j-1}). 
\label{action-BCS0}
\end{equation}
It is important to notice here that the phase space for the path ${\bf z}$
is the complex plane ${\bf C}$, which is the universally covering space for
the cylinder $\Gamma$. 
The transition amplitude has been represented as a sum
of transition amplitudes each of which is associated with 
the final state
$z_{{\rm F}}^{m}=z_{{\rm F}}+i2\pi m$ with a specified $m$. Note that
$z_{{\rm F}}^{m}=
\kappa x_{{\rm F}}+i\phi_{{\rm F}}^{m}$ with $\phi_{{\rm F}}^{m} :=
\phi_{{\rm F}}+2\pi m$.
Although $|z_{{\rm F}}^{m}\rangle$ and $|z_{{\rm F}}\rangle$ are an
identical state of
the Bloch particle,
one need to distinguish them and sum over the relevant paths in the path
integral formula. 
This is derived from that $\Gamma$ is not simply connected \cite{Schulman}. 

We shall rewrite the expression for ${\cal S}[{\bf z}^{*},{\bf z}]$ into a
more tractable form.
It is composed of the canonical term and the dynamical term: 
\begin{eqnarray}
{\cal S}[{\bf z}^{*},{\bf z}] &=& {\cal S}_{{\rm c}}[{\bf z}^{*},{\bf z}] +
{\cal S}_{{\rm d}}[{\bf z}^{*},{\bf z}], \label{action-BCS}\\
{\cal S}_{{\rm c}}[{\bf z}^{*},{\bf z}] &:=& \sum_{j=1}^{N}{\cal S}_{{\rm
c}}(z_{j}, z_{j-1}), \label{action-BCS2}\\
{\cal S}_{{\rm d}}[{\bf z}^{*},{\bf z}] &:=&
-\sum_{j=1}^{N}\epsilon\,H(z_{j}, z_{j-1}). \label{action-BCS3}
\end{eqnarray}
We introduce here a new quantity by the equation
\begin{equation}
\frac{i}{\hbar}{\cal S}_{{\rm c}}(z_{j}, z_{j-1}) = 
P(z_{j}) - P(z_{j-1}) + 
\frac{i}{\hbar}{\tilde {\cal S}}_{{\rm c}}(z_{j}, z_{j-1})
\end{equation}
with 
\begin{equation}
P(z) := \frac{1}{8\kappa a}\left\{(z^{*})^{2} - z^{2}\right\}. 
\end{equation}
Then, we may write
\begin{equation}
\frac{i}{\hbar}{\cal S}_{{\rm c}}[{\bf z}^{*},{\bf z}] = 
P(z_{{\rm F}}^{m}) - P(z_{{\rm I}}) + 
\frac{i}{\hbar}{\tilde {\cal S}}_{{\rm c}}[{\bf z}^{*},{\bf z}], 
\label{action-BCS4}
\end{equation}
\begin{equation}
\frac{i}{\hbar}{\tilde {\cal S}}_{{\rm c}}[{\bf z}^{*},{\bf z}] := 
\sum_{j=1}^{N}\left(-\frac{1}{4\kappa a}\right)\left\{
z^{*}_{j}(z_{j} - z_{j-1}) - (z^{*}_{j} - z^{*}_{j-1})z_{j-1} 
\right\},  
\label{action-BCS5}
\end{equation}
where the summand is just an explicit form for $(i/\hbar){\tilde {\cal
S}}_{{\rm c}}(z_{j}, z_{j-1})$.
The first two term
 in the r.\ h.\ s.\ of (\ref{action-BCS}) are purely imaginary, so that
they may be called gauge terms because they give rise to a phase factor of
(\ref{T-amplitude}). The phase factors affect the interference among
different terms in (\ref{summ-amplitude}).
 
The equation (\ref{action-BCS5}) reduces in the continuum-time limit to the
functional:
\begin{equation}
\frac{i}{\hbar}{\tilde {\cal S}}_{{\rm c}}[z^{*}, z] = 
-\frac{1}{4\kappa a}\int_{0}^{T}dt\,
\left\{z^{*}(t)\frac{dz(t)}{dt} - \frac{dz^{*}(t)}{dt}z(t)\right\}.
%-\frac{i}{\hbar}\int_{0}^{T}dt\,
%H(z(t), z(t)).
\label{continuum-limit}
\end{equation}
%\seeqa
However, 
we should emphasize on the necessity of the discrete-time formalism 
of the coherent state path integral in order to obtain a correct 
result for the transition amplitude \cite{Shibata-Takagi1}. 
Thus, equation (\ref{continuum-limit}) is usable only for the evaluation of
the stationary action,
We should mention, finally, that the discrete-time formalism 
of the coherent state path integral can be consistently formulated 
because ${\tilde {\cal S}}_{{\rm c}}[{\bf z}^{*},{\bf z}]$ 
and ${\cal S}_{{\rm d}}[{\bf z}^{*},{\bf z}]$ are holonomic 
functions of two sets of complex variables ${\bf z}$ and ${\bf z}^{*}$, 
where the two sets are regarded to be independent variables
\cite{Shibata-Takagi1}.

%%%%%%%%%%%%%%%%%%%%%%%%%%%%%%%%%%%%%%%%%%
\section{Application to the Bloch Oscillation}
%%%%%%%%%%%%%%%%%%%%%%%%%%%%%%%%%%%%%%%%%%%

As an application of the present formalism, 
we consider the time evolution of the coherent state of the Bloch particle 
under a uniform electric field $F$. 
As a consequence of the boundedness of the energy band, 
the Bloch particle exhibits so called the Bloch oscillation 
\cite{Bloch2}, \cite{Zener}. 
If the simplest dispersion (\ref{simp-disp}) is adopted, the Hamiltonian 
assumes %#### 
\begin{equation}
\hat{H} = -V\cos \hat{\phi} - F \hat{x}, 
\end{equation}
whose matrix element is: 
\begin{equation}
H(z_{j}^{*},z_{j-1})
=-Ve^{-\kappa a/4}\cosh\left(\frac{z^{*}_{j}-z_{j-1}}{2}\right) 
-\frac{F}{2\kappa}(z^{*}_{j}+z_{j-1}). 
\end{equation}
To evaluate the transition amplitude, 
we employ the stationary-action approximation, 
which is justified for a semi-classical motion.

We consider here the classical motion of the Bloch particle. The equation
of motion is:
\begin{eqnarray}
&&\frac{dx}{dt}=\frac{aV}{\hbar}\sin \phi,\\
&&\frac{d\phi}{dt}=\omega
\label{cl-eq-motion}
\end{eqnarray}
with
\begin{equation}
\omega:= \frac{Fa}{\hbar}.
%,\qquad \Omega := \frac{\kappa a}{\hbar}Ve^{-\kappa a/4}.
\end{equation}
The solution for the initial condition, $x(0) = x_{{\rm I}}$ and $\phi(0) =
\phi_{{\rm I}}$, is given by 
\begin{eqnarray}
&&x = x_{\rm cl.}(t) := x_{{\rm I}} + L_{\rm cl.}[\cos{\phi_{{\rm I}}} -
\cos{(\omega t
+ \phi_{{\rm I}}})], \\
&&\phi = \phi_{\rm cl.}(t) := \phi_{{\rm I}} + \omega t, 
\label{classical motion}
\end{eqnarray} 
where $L_{\rm cl.} := V/F$ 
is so called the localization length. This solution shows the Bloch
oscillation with the angular frequency $\omega$.

Thus, the condition for the motion to be semi-classical is given by %####
$L_{\rm cl.} \gg \delta x \simeq \sqrt{a/\kappa} \gg a$. %####

%%%%%%%%%%%%%%%%%%%%%%%%%%%%%%%%%%
\subsection{Stationary action}
%%%%%%%%%%%%%%%%%%%%%%%%%%%%%%%%%%

The primary task is to find the stationary point, namely, stationary action
path 
of the action ${\cal S}[{\bf z}^{*},{\bf z}]$ 
or, equivalently, ${\tilde {\cal S}}[{\bf z}^{*},{\bf z}]$. %####
We will execute it by a
similar procedure to the one presented in Ref.\cite{Shibata-Takagi1}. 
The equation of motion for the stationary action path is 
obtained by differentiating the action:
\begin{equation}
\left.\frac{\partial{\tilde {\cal S}}[{\bf z}^{*},{\bf z}]}{\partial
z^{*}_{j}}%####
\right|_{{\rm s}}=
\left.\frac{\partial{\tilde {\cal S}}[{\bf z}^{*},{\bf z}]}{\partial z_{j}}%####
\right|_{{\rm s}}=0,\qquad j=1,2,\cdots,N-1.
\end{equation}
For the stationary action path $\{{\bf z}^s,{\bar {\bf z}}^s\}$, 
we are allowed to take the continuous-time limit \cite{Shibata-Takagi1}, so
that, 
%\sbeqa
\begin{eqnarray}
&&\frac{dz^{{\rm s}}(t)}{dt}=i\omega 
+ i\Omega\sinh\left(\frac{{\bar z}^{{\rm s}}(t)-z^{{\rm s}}(t)}{2}\right),\\
&&\frac{d{\bar z}^{{\rm s}}(t)}{dt}=-i\omega 
+ i\Omega\sinh\left(\frac{{\bar z}^{{\rm s}}(t)-z^{{\rm s}}(t)}{2}\right)
\label{eq-motion}
\end{eqnarray}
%\seeqa
with
\begin{equation}
\omega:= \frac{Fa}{\hbar},\qquad
\Omega := \frac{\kappa a}{\hbar}Ve^{-\kappa a/4}.
\end{equation}
These equations must be solved under the following boundary conditions
\cite{Shibata-Takagi1}:
\begin{equation}
z^{{\rm s}}(0)=z_{{\rm I}},\qquad {\bar z}^{{\rm s}}(T)=z_{{\rm F}}^{m*}.
\label{boundary conditions}
\end{equation}
It should be noted that $z^{{\rm s}}(t)$ and ${\bar z}^{{\rm s}}(t)$ are
not necessary
complex conjugate to each other \cite{Shibata-Takagi1}.

From (\ref{eq-motion}) we obtain
\begin{equation}
\frac{1}{2}(z^s(t) - {\bar z}^s(t)) = i(\omega t+\phi_{0}),
\label{delta-z}
\end{equation}
where $\phi_{0}$ is a complex constant. Inserting this into the r.\ h.\
s\verb/./'s of (\ref{eq-motion}), we can solve for $z^s(t)$ and ${\bar
z}^s(t)$ to obtain
\begin{eqnarray}
&&z^s(t) = z^{{\rm s}}(t, \phi_{0}) := z_{I}+i\omega t -\frac{\Omega}{\omega}
\left[\cos(\omega t + \phi_{0})-\cos\phi_{0}\right], \label{sapa} \\
&&{\bar z}^s(t) = {\bar z}^{{\rm s}}(t, \phi_{0}) := z_{F}^{m*}-i\omega
(t-T) -\frac{\Omega}{\omega}
\left[\cos(\omega t + \phi_{0})-\cos(\omega T+\phi_{0})\right],
\label{sapb} \end{eqnarray}
%\seeqa
where the integration constants are fixed by the boundary conditions
(\ref{boundary conditions}). 
This solution is consistent to (\ref{delta-z}) only if %####
$\phi_{0}$ satisfies the equation, 
\begin{equation}
D(\phi_{0}) := z_{F}^{m*} - z^s(T, \phi_{0}) + 2i(\omega T + \phi_{0}) = 0, 
\label{alge1}
\end{equation}
where use has been made of ${\bar z}^s(T, \phi_{0}) = z_{{\rm F}}^{m*}$.

The following two quantities are introduced here for a convenience of later
arguments:
\begin{equation}
{\tilde \Omega}(t) := \frac{\Omega}{2}\cos (\omega t + \phi_{0}), 
\label{tilde-Omega}
\end{equation}
\begin{equation}
Z(t, \phi_{0}) := \int_{0}^{t}dt'{\tilde \Omega}(t') = 
\frac{\Omega}{2\omega} [\sin{(\omega t + \phi_{0})} - \sin{\phi_{0}}].
\label{Z(t)}
\end{equation}
Alternatively, we may write 
\begin{equation}
Z(t, \phi_{0}) = \frac{1}{2}\frac{\partial{z^s(t, \phi_{0})}}{\partial
\phi_{0}}.
\label{Z(t)-2}
\end{equation}

The stationary action, %####
${\cal S}^{{\rm s}}_{m} := {\cal S}[{\bar {\bf z}}^s,{\bf z}^s]$, %####
can be obtained by %####
substituting (\ref{sapa}) and (\ref{sapb}) into the action
(\ref{action-BCS}) %####
with (\ref{action-BCS3}), (\ref{action-BCS4}), and (\ref{continuum-limit}):%####
\begin{eqnarray}
&&\frac{i}{\hbar}{\cal S}^{{\rm s}}_{m}
=-\frac{i}{2a}\left\{
x_{{\rm F}}\phi_{{\rm F}}^{m} - x_{{\rm I}}\phi_{{\rm I}}
\right\}\nonumber \\
&&-\frac{1}{4\kappa a}\bigg[
|z_{{\rm F}}^{m}|^{2}+|z_{{\rm I}}|^{2}-2z_{{\rm F}}^{m*}z_{{\rm I}}
-(z_{{\rm F}}^{m*}-z_{{\rm I}}+i\omega T + 2i\phi_{0})^{2}\nonumber \\
&&
-2i\omega T\left(z_{{\rm F}}^{m*}+z_{{\rm I}} + i\frac{\omega T}{2}\right)
-8iZ(T, \phi_{0})
\bigg].
\label{stationary-action}
\end{eqnarray}

%%%%%%%%%%%%%%%%%%%%%%%%%%%%%%%%%%
\subsection{Semi-classical stationary action} 
%%%%%%%%%%%%%%%%%%%%%%%%%%%%%%%%%%

Let $z_{{\rm sc}}(t) := z^{{\rm s}}(t, \phi_{I})$. Then $z(t) := z_{{\rm
sc}}(t)$ and ${\bar z}(t) := z_{{\rm sc}}^{*}(t)$
satisfy the equation of motion (\ref{eq-motion}) 
with the initial condition, $z(0) = z_{{\rm I}}$ 
and ${\bar z}(0) = z_{{\rm I}}^{*}$. More explicitly, 
\begin{eqnarray}
&&z_{\rm sc.}(t) = \kappa x_{\rm sc.}(t) + i\phi_{\rm sc.}(t),
\label{xclassical motion} \\
&&x_{\rm sc.}(t) = x_{{\rm I}} + L[\cos{\phi_{{\rm I}}} - \cos{(\omega t +
\phi_{{\rm I}}})], \\
&&\phi_{\rm sc.}(t) = \phi_{{\rm I}} + \omega t
\end{eqnarray} 
with 
\begin{equation}
L := \frac{\Omega}{\omega\kappa} = \frac{V}{F}e^{-\kappa a/4}.
%,\qquad \Omega := \frac{\kappa a}{\hbar}Ve^{-\kappa a/4}.
\end{equation}
The solution (\ref{xclassical motion}) is the semi-classical solution,
which tends to the classical solution (\ref{classical motion}) in the
limit $\kappa a \rightarrow 0$ because $L = L_{\rm cl.}\exp{(-\kappa
a/4)}$. 

Let $\Delta z := z_{{\rm F}}^{m} - z_{{\rm sc}}(T)$. Then 
\begin{equation}
\Delta z = \kappa \Delta x + i\Delta \phi, 
\end{equation}
\begin{equation}
\Delta x = x_{{\rm F}}-x_{\rm sc.}(T), 
\end{equation}
\begin{equation}
\Delta \phi = \phi_{{\rm F}}^{m}-\phi_{\rm sc.}(T).
\end{equation}
The condition for the semi-classical solution to satisfy the boundary
conditions (\ref{boundary conditions}) is given by $\Delta z = 0$ 
for an integer $m$. If this condition is satisfied, the
solution of the equation (\ref{alge1}) is given by $\phi_{0} = \phi_{{\rm I}}$.

It will be shown later on that under the semi-classical condition, %####
we can assume %####
$|\Delta z|^2 = (\kappa\Delta x)^2 + (\Delta \phi)^2 \ll 1$. %####
Then, we are allowed to solve the equation (\ref{alge1}) for $\phi_{0}$ to
the first order on $\Delta z$. 
We set $\phi_{0} := \phi_{{\rm I}} + \delta \phi$, and 
expand $D(\phi_{0})$ to the first order on $\delta \phi$ to obtain
\begin{equation}
D(\phi_{0}) \simeq \Delta z^{*} + 2(i - Z(T,\phi_{\rm I}))\delta \phi, 
\end{equation}
where use has been made of 
$D(\phi_{\rm I}) = \Delta z^{*}$ and (\ref{Z(t)-2}).
Therefore, we obtain
\begin{equation}
\phi_{0} \simeq 
\phi_{{\rm I}} + \frac{i\Delta z^{*}}{2\{1 + iZ(T,\phi_{\rm I})\}}. 
\label{phi0}
\end{equation}
In the similar way, let 
$\tilde{z}_{{\rm sc}}(t):=z^{{\rm s}}(t-T,\phi_{{\rm F}}^{m})$ and 
$\Delta\tilde{z}_{{\rm sc}}(0)-z_{{\rm I}}$. 
Then, under the semi-classical condition, 
$|\Delta\tilde{z}|^{2}\ll 1$, 
we can obtain the correspondence with  $\delta \phi$ as 
\begin{equation}
\delta\tilde{\phi}:= \frac{i\Delta\tilde{z}}{2\{1+iZ(T,\phi_{{\rm F}}^{m}-\omega T)\}}. 
\end{equation}
We should remark here that the following equality holds to the zeroth
order on $\Delta \phi$: 
\begin{equation}
Z(T,\phi_{{\rm I}})
\simeq
Z(T,\phi_{{\rm F}}^{m}-\omega T)
\simeq \frac{\Omega}{\omega}
\cos\left(\frac{\phi_{{\rm F}}^{m} + \phi_{{\rm
I}}}{2}\right)\sin\frac{\omega T}{2}
=: C(T), \label{C(T)}
\end{equation}
which is symmetrical with respect to $\phi_{{\rm I}}$ and $\phi_{{\rm F}}^{m}$. Hence, we set 
\begin{eqnarray}
&&\phi_{0}=\frac{\phi_{{\rm F}}^{m}+\phi_{{\rm I}}-\omega T }{2}
+\frac{i\kappa\Delta\bar{x}}{2\{1+iC(T)\}},
\label{phi0}\\
&&\Delta\bar{x}:=x_{{\rm F}}-x_{{\rm I}}
+ L\left[\cos\left(\frac{\phi_{\rm F}+\phi_{\rm I}+\omega T}{2}\right)
-\cos\left(\frac{\phi_{\rm F}+\phi_{\rm I}-\omega T}{2}\right)\right], 
\end{eqnarray}
and substituting (\ref{phi0}) into 
(\ref{stationary-action}) and (\ref{prefactor}), 
we find the following semi-classical stationary action: 
\begin{equation}
\frac{i}{\hbar}{\cal S}^{{\rm s}}_{m} = 
-\frac{\kappa}{4a}\frac{(\Delta {\bar x})^{2}}{1 + iC(T)} - \frac{1}{4\kappa
a}(\Delta \phi)^{2} 
-\frac{i}{2a}(x_{{\rm F}}+x_{{\rm I}})\Delta \phi + \frac{2i}{\kappa a}C(T)
\label{semi-classical-action}
\end{equation}

%%%%%%%%%%%%%%%%%%%%%%%%%%%%%%%%%%
\subsection{Fluctuation} 
%%%%%%%%%%%%%%%%%%%%%%%%%%%%%%%%%%

We introduce the fluctuation variables through 
\begin{equation}
z_{j} = z^{{\rm s}}_{j}+\zeta_{j},\qquad
z^{*}_{j} = {\bar z}^{{\rm s}}_{j} + \zeta^{*}_{j},
\qquad j=1,2,\cdots,N-1. 
\label{fluc-BCS}
\end{equation}
%with $\zeta(0)=\zeta^{*}(N)=0$. 
Substituting (\ref{fluc-BCS}) into the action ${\cal S}[{\bar {\bf z}},
{\bf z}]$, %####
and expanding the result up to the second order in the
fluctuation, we obtain 
%\sbeqa
\begin{equation}
{\cal S}[{\bar {\bf z}}^s+\zeta^{*},{\bf z}^s+\zeta]
\simeq
{\cal S}^{{\rm s}}_{m}+{\cal S}^{2}_{m}[\zeta^{*},\zeta],
\label{fluc-action}
\end{equation}
where ${\cal S}^{2}_{m}[\zeta^{*},\zeta]$ is a quadratic form on the
fluctuation variables. Therefore, the contribution of the fluctuation to
the action is represented as a factor of
\begin{equation}
{\cal F}_{m}(T):=\lim_{N\to\infty}
\int_{-\infty}^{\infty}\prod_{j=1}^{N-1}
\frac{d(\Re\zeta_{j})d(\Im\zeta_{j})}{2\kappa a}
\exp\left(\frac{i}{\hbar}{\cal S}^{2}_{m}[\zeta^{*},\zeta]\right), 
\end{equation}
which is a multi-dimensional Gaussian integral. The leading contribution to
$(i/\hbar){\cal S}^{2}_{m}[\zeta^{*},\zeta]$ is written as
$-\sum_{j}\zeta^{*}_{j}\zeta_{j}/(2\kappa a)$, and the fluctuation of %####
$\zeta$ is of the order $\sqrt{\kappa a}$, which is small %####
when $\kappa a \ll 1$. %####

There exists an standard procedure to evaluate ${\cal F}_{m}(T)$ 
\cite{Solari}, \cite{Shibata-Takagi2}. %####
To evaluate the Gaussian integral is reduced to a calculation of the
determinant of a symmetric matrix associated with the quadratic form. Since
the matrix is tridiagonal its determinant is given as the last term of
coupled three-term recursion relations. 
In the limit of $\epsilon \to 0$, 
they reduce to a set of coupled first-order differential equations:
%\sbeqa
\begin{eqnarray}
&&\frac{dM(t)}{dt} = -{\tilde \Omega}(t)M'(t), \\
&&\frac{dM'(t)}{dt} = -{\tilde \Omega}(t)\left\{M(t) + 2iM'(t)\right\} \\
\end{eqnarray}
with the initial condition 
\begin{equation}
M(0)=1,\qquad M'(0)=0.
\end{equation}
The differential equations are readily solved to obtain
%\sbeqa 
\begin{eqnarray}
&&M(t) = e^{-iZ(t)}\left[1+iZ(t)\right],\\ 
&&M'(t) = -Z(t)\,e^{-iZ(t)} 
\end{eqnarray}
with $Z(t) := Z(t, \phi_{0})$. Hence, %####
\begin{eqnarray}
{\cal F}_{m}(T) &=& \left[M(T)\right]^{-1/2} \\
&=& \frac{e^{iC(T)/2}}{\sqrt{1+iC(T)}}, 
\label{prefactor}
\end{eqnarray}
where we have used the semi-classical expression, $C(T) \simeq Z(T,
\phi_{I})$. %####

%%%%%%%%%%%%%%%%%%%%%%%%%%%%%%%%%%
\subsection{Transition amplitude}
%%%%%%%%%%%%%%%%%%%%%%%%%%%%%%%%%%

The transition amplitude in the semi-classical approximation is given by
\begin{eqnarray}
&&{\cal K}(z_{{\rm F}},z_{{\rm I}};T)\simeq\sum_{m=-\infty}^{\infty}
{\cal L}^{{\rm s}}(z_{{\rm F}}^{m},z_{{\rm I}};T), \label{K-amplitude} \\ 
&&{\cal L}^{{\rm s}}(z_{{\rm F}}^{m},z_{{\rm I}};T):= 
{\cal F}_{m}(T)\,e^{i{\cal S}^{{\rm s}}_{m}/\hbar}, \label{L-amplitude} \\
&&{\cal F}_{m}(T)=\frac{1}{\sqrt{1+iC(T)}}.
\end{eqnarray}
It follows that
\begin{equation}
|{\cal L}^{{\rm s}}(z_{{\rm F}}^{m},z_{{\rm I}};T)|^2 = 
\frac{1}{\sqrt{1+(C(T))^2}}\exp{\left[-\frac{\kappa}{2a}\frac{(\Delta
{\bar x})^{2}}{1 + (C(T))^2} - \frac{1}{2\kappa a}(\Delta \phi)^{2}\right]} 
\label{probability}
\end{equation}
or, equivalently, 
\begin{eqnarray}
&&|{\cal L}^{{\rm s}}(z_{{\rm F}}^{m},z_{{\rm I}};T)|^2 = 
\frac{1}{\sqrt{1+(C(T))^2}}\exp{\left[-\frac{\kappa}{2a}\frac{(x_{{\rm
F}}-{\bar x}_{\rm
sc.}(T))^{2}}{1 + (C(T))^2} - \frac{1}{2\kappa a}(\phi_{{\rm F}}^{m}-\phi_{\rm
sc.}(T))^{2}\right]}
\label{probability2}, \\
&&{\bar x}_{{\rm sc}}(T)
:= x_{{\rm I}}+ 2L\sin\left(\frac{\phi_{\rm F}^{m}+\phi_{\rm I}}{2}\right)
\sin\frac{\omega T}{2}. 
\end{eqnarray}
For given $x_{{\rm F}}$ and $\phi_{{\rm F}}$, the summation in the r.\ h.\
s.\ of
(\ref{K-amplitude}) is dominated by a single term because we have assumed
that $a\kappa \ll 1$. Therefore, the interference among terms with
different winding numbers vanishes then. Since $\phi_{\rm sc.}(T)$ changes
linearly with $T$, the leading term in (\ref{K-amplitude}) changes
discontinuously with $T$.
%the value of $m$ of the leading term in (\ref{K-amplitude}) increases
%stepwise with $T$.
It follows from (\ref{probability2}) that, for a given initial condition
($x_{{\rm I}}, \phi_{{\rm I}}$), 
the wave packet centroid in the extended phase space,
$x_{{\rm F}}$-$\phi_{{\rm F}}$, exhibits the Bloch oscillations along the
position axis.
While, the time-dependence of the spreads of the wave packet in the phase
space are given by
\begin{equation}
\delta x (T)= \sqrt{\frac{2a}{\kappa}\{1 + (C(T))^2\}}, \qquad %####
\delta\phi=\sqrt{2\kappa a} 
\label{delta x-phi}
\end{equation}
along the momentum and the position axes, respectively. The maximum value
of %####
$\delta x (T)$ is equal to $\sqrt{(2a/\kappa)\{1 + (C(T))^2\}}$, so that %####
our assumption $|\Delta z|^2 \ll 1$ has been justified under the
semi-classical condition. %####

The expressions for
$\delta x(0)$ and $\delta\phi$ are larger by the factor $\sqrt{2}$ %####
than those for the relevant coherent states. This is because $\langle
z_{{\rm F}}^{m}|z_{{\rm I}} \rangle$ is a sort of the convolution of the
two coherent
states.
The time-dependence of the spread of the wave packet along the position
axis is periodic by (\ref{delta x-phi}) with (\ref{C(T)}). This breathing
behavior is similar to the behavior of the squeezed state of the harmonic
oscillator \cite{Walls-Milburm}.

Our solution remains correct in the zero-field limit, $F = 0$, where the
Bloch particle exhibits a ballistic motion: $x_{\rm sc.}(T) = x_{{\rm I}} +
v_{{\rm I}}T$, $\phi_{\rm sc.}(T) = \phi_{{\rm I}}$, and $C(T) = (\Omega
T/2)\cos
{\phi_{{\rm I}}}$ with $v_{{\rm I}} := (aV/\hbar)\sin {\phi_{{\rm I}}}$
being the group
velocity. The spread of the wave packet, $\delta x(T)$, increases monotonously
with $T$ because of the dispersion of the phase velocity.

%%%%%%%%%%%%%%%%%%%%%%%%%%%%%%%%%%%%%%%%%%
\section{Summary and Discussion}
%%%%%%%%%%%%%%%%%%%%%%%%%%%%%%%%%%%%%%%%%%

We have succeeded in properly defining the coherent states for 
the Bloch particle. Using them,  
we have constructed a coherent state path integral formalism for 
the Bloch particle. 
The transition amplitude involves contribution from paths with different
winding numbers 
associated with the quasi-momentum. 
The theory has been successfully applied to the semiclassical motion of the
Bloch particle 
under a uniform electric field, and the famous Bloch oscillation has been %####
reproduced.

If one of two dynamical variables which are canonically conjugate to each
other is continuous but periodic, the other is discrete and unbounded on
both the positive and the negative sides of its value. The converse of this
statement is also true. The second example in addition to the Bloch
particle is the rigid rotator, where the rotation angle, $\phi$, is
continuous but the angular-momentum, $L_z$, is discrete. That is, the roles
%####
of the position and the momentum are reversed from the case of the Bloch
particle. Therefore, the roles of the kinetic energy and the potential %####
energy are reversed as well. The periodic potential energy $W(\phi)$ stands
for the hindering potential, and can assume various forms depending on the
physical condition for the rigid rotator. On the contrary, the kinetic %####
energy is restricted to the form $(L_z)^2/(2I)$ with $I$ being the moment
of inertia. 
The third example is the order parameter of a mesoscopic superconductor or
the BEC of a ultra cold gas, where the phase of the order parameter,
$\phi$, is continuous but the number of the condensate, $N$, is discrete.
Anyway, the coherent state path integral formalism established in the
present paper can be applied to other systems than the Bloch particle. 

It is our hope that the path integral formalism will 
help to achieve solving more complicated problem 
for the Bloch particle or other systems. 
In fact, the authors have found recently equivalence of the quantum
dynamics of a 
domain wall in a quasi-one dimensional mesoscopic 
ferromagnet to that of the Bloch particle. 
This subject is discussed in a separate paper \cite{Shibata-Niizeki}. %####

%%%%%%%%%%%%%%%%%%%% References %%%%%%%%%%%%%%%%%%%%%
%%%%%%%%%%%%%%%%%%
%\begin{references}
%%%%%%%%%%%%%%%%%%

\end{document}